\documentclass[pre]{revtex4}

\usepackage{amsmath}    
\usepackage{amsfonts}
\usepackage{amssymb}
\usepackage{graphicx}   
\usepackage{dcolumn}
\usepackage{bm}

\begin{document}

\title{Filamentation instability in a quantum magnetized plasma}

\author{A. Bret}
\email{antoineclaude.bret@uclm.es}
 \affiliation{ETSI Industriales, Universidad de Castilla-La Mancha, 13071 Ciudad Real, Spain}
 \affiliation{Instituto de Investigaciones Energ\'{e}ticas y Aplicaciones Industriales, Campus Universitario de Ciudad Real, 13071 Ciudad Real, Spain}

\date{\today }

\begin{abstract}
The filamentation instability occurring when a non relativistic electron beam passes through a quantum magnetized plasma is investigated by means of a cold quantum magnetohydrodynamic model. It is proved that the instability can be completely suppressed by quantum effects if and only if a finite magnetic field is present. A dimensionless parameter is identified which measures the strength of quantum effects. Strong quantum effects allow for a much smaller magnetic field to suppress the instability than in the classical regime.
\end{abstract}

\maketitle

\section{Introduction}
The  development of quantum hydrodynamic and magnetohydrodynamic equations \cite{Gardner96,Haas2005} made it possible to quickly evaluate quantum effects connected to the physics of microelectronic devices and laser plasmas interaction (see Ref. \cite{Ren2007} and references therein). Plasma physics has also gained from these progresses as quantum effects appear in Fusion settings or Astrophysics. The behavior of waves in quantum plasmas \cite{HaasGarcia2003,ShuklaPhysLettA2006,ShuklaJPP2006,ShuklaPoP2006,Ren2007} magnetized or not, as well as turbulence is such environments \cite{Shaikh2007} has thus received attention. Another very classical topic of plasma physics, namely plasma instabilities, needs to be revisited from the quantum point of view. The quantum theory of the two-stream instability has already been developed \cite{Haas2000,Anderson2002} while quantum effects on the  filamentation instability were recently evaluated \cite{BretPoPQuantum2007} for a non magnetized plasma. Due to the importance of magnetized plasmas, especially in astrophysics, we devote the present paper to the evaluation of quantum effects on the filamentation instability in such setting. Since the relativistic quantum magnetohydrodynamic equations are yet to be defined, the present analysis restricts to the non-relativistic regime. On the other hand, we do not make any approximation on the beam density so that present theory remains valid even when the beam density equals the plasma electronic one.

The paper is structured as follow: we start explaining the formalism and derive the  dispersion equation. We then turn to the investigation of the marginal stability and derive some exact relations satisfied in this case. We finally study the maximum growth rate and the most unstable wave vector before we reach our conclusions.

Let us then consider an infinite and homogenous cold non-relativistic
electron beam of velocity $V_b\mathbf{z}$ and density $n_b$ entering a cold plasma along the guiding magnetic field $\mathbf{B}_0=B_0\mathbf{z}$.
The plasma has the electronic
density $n_p$ and ions form a fixed neutralizing background of
density $n_b+n_p$. The beam prompts a return current in the
plasma with velocity $V_p\mathbf{z}$ such as
$n_p\mathbf{V}_p=n_b\mathbf{V}_b$. We use the fluid conservation equations
for the beam ($j=b$) and the plasma ($j=p$),
\begin{equation}\label{eq:conser}
   \frac{\partial n_j}{\partial t}+\nabla\cdot(n_j \mathbf{v}_j)=0
\end{equation}
and the force equation in the presence of the static magnetic field $\mathbf{B}_0$ with a Bohm potential term \cite{Haas2005},
\begin{equation}\label{eq:force}
   \frac{\partial \mathbf{v}_j}{\partial
   t}+(\mathbf{v}_j\cdot\nabla)\mathbf{v}_j=-\frac{q}{m}\left(\mathbf{E} + \frac{\mathbf{v}_j\times
   \mathbf{B}}{c}\right)+\frac{\hbar^2}{2
   m^2}\nabla\left(\frac{\nabla^2\sqrt{n_j}}{\sqrt{n_j}}\right),
\end{equation}
where $q>0$ and $m$ are the charge and mass of the electron, $n_j$
the density of species $j$, $p_j$ its momentum, and $\mathbf{B}$ equals $\mathbf{B}_0$ plus the induced magnetic field. We now study the response of the system to density perturbations with $\mathbf{k}\perp\mathbf{V}_b$, varying like $\exp(\imath \mathbf{k}\mathbf{r}-\omega t)$ with $\mathbf{k}=k\mathbf{x}$, and linearize the equations above. With the subscripts 0
and 1 denoting the equilibrium and perturbed quantities
respectively, the linearized conservation equation
(\ref{eq:conser}) yields
\begin{equation}\label{eq:conserL}
  n_{j1} = n_{j0} \frac{\mathbf{k}\cdot \mathbf{v}_{j1}}{\omega -\mathbf{k}\cdot
  \mathbf{v}_{j0}},
\end{equation}
and the force equation (\ref{eq:force}) gives
\begin{equation}\label{eq:forceL}
   i (\mathbf{k}\cdot \mathbf{v}_{j0}-\omega)\mathbf{v}_{j1} \nonumber\\
   = -\frac{q}{m}\left(\mathbf{E}_{1}+\frac{\mathbf{v}_{j0}\times\mathbf{B}_1+\mathbf{v}_{j1}\times\mathbf{B}_0}{c}\right)
   -i\frac{\hbar k^2}{4m^2}\frac{n_{j1}}{n_{j0}}\mathbf{k}.
\end{equation}
From the linearized equations above, we derive the perturbed density and velocity fields in terms of $\mathbf{E}_{1}$ and $\mathbf{B}_1$ and eventually express the current through,
\begin{equation}\label{eq:current}
   \mathbf{J} = q\sum_{j=p,b} n_{j0}\mathbf{v}_{j1}+n_{j1}\mathbf{v}_{j0}.
\end{equation}
Finally, we express $\mathbf{B}_1$ in terms of $\mathbf{E}_1$ through $\mathbf{B}_1=(c/\omega)\mathbf{k}\times\mathbf{E}_1$ and close the system inserting the current expression in a combination of Maxwell Amp\`{e}re and Faraday's equations,
\begin{equation}\label{eq:Maxwell}
  \frac{c^2}{\omega^2}\mathbf{k}\times(\mathbf{k}\times \mathbf{E}_1)+\mathbf{E}_1 + \frac{4i\pi}{\omega}\mathbf{J} = 0
  \Leftrightarrow \mathbf{T}(\mathbf{E}_1)=0.
\end{equation}
The tensor $\mathbf{T}$ has here been calculated symbolically using an adapted version of the \emph{Mathematica} Notebook described in Ref. \cite{BretCPC}. It takes the form
\begin{equation}\label{eq:tensor}
    \mathbf{T} =\left(%
\begin{array}{ccc}
  T_{11} & T_{12}^* & 0 \\
  T_{12} & T_{22} & 0 \\
  0 & 0 & T_{33} \\
\end{array}%
\right),
\end{equation}
where the superscript * refers to the complex conjugate and
\begin{eqnarray}\label{eq:T11}
    T_{11}&=&x^2\left(1-\frac{(1+\alpha)}{(x^2-\Omega_B^2)-\Theta Z^4}\right),\nonumber\\
    T_{22}&=&x^2-\frac{Z^2}{\beta^2}-\frac{(1+\alpha)(\Theta Z^4-x^2)}{(x^2-\Omega_B^2)-\Theta Z^4},\nonumber\\
    T_{33}&=&x^2-1-\alpha-\frac{Z^2}{\beta^2}\left(1+\frac{\alpha(1+\alpha)\beta^2}{(x^2-\Omega_B^2)-\Theta Z^4}\right),\nonumber\\
    T_{12}&=&\imath \frac{x(1+\alpha)\Omega_B}{(x^2-\Omega_B^2)-\Theta Z^4},
\end{eqnarray}
in terms of
\begin{equation}\label{eq:param}
    x=\frac{\omega}{\omega_p},~Z=\frac{kV_b}{\omega_p},~\beta=\frac{V_b}{c},~\alpha=\frac{n_b}{n_p},~\Omega_B=\frac{qB_0}{mc\omega_p},
\end{equation}
where $\omega_p$ is the electronic plasmas frequency. Quantum effects appear to be measured through a parameter previously highlighted \cite{Haas2000,Anderson2002,BretPoPQuantum2007},
\begin{equation}\label{eq:theta}
    \Theta=\frac{\Theta_c}{\beta^4},~~\mathrm{with}~~\Theta_c=\left(\frac{\hbar\omega_p}{2mc^2}\right)^2 .
\end{equation}
Numerically,
\begin{equation}\label{eq:ThetaNum}
    \Theta_c=1.3\times 10^{-33}n_p~~[\mathrm{cm}^{-3}],
\end{equation}
so that this parameter will hardly be larger than 1, even when dealing with the densest space plasmas.

\begin{figure}[t]
\begin{center}
\includegraphics[width=0.45\textwidth]{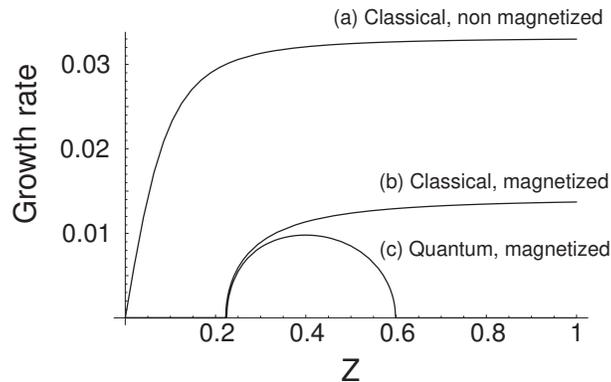}
\end{center}
\caption{Classical (non-quantum) growth rate of the filamentation instability in terms of the reduced wave vector $Z$ without (a) and with (b) magnetic field. Curve (c) includes quantum effects. Parameters are $\alpha=0.1$, $\beta=0.1$ for (a,b,c), $\Omega_B=0.03$ for (a,b) and $\Theta_c=1.3\times 10^{-7}$ for (c), which corresponds to the plasma density $n_p=10^{26}$ cm$^{-3}$.} \label{fig:1}
\end{figure}

Let us finally emphasized a point regarding the beam to plasma density ratio $\alpha$ defined in Eq. (\ref{eq:param}). If the ground state which stability is investigated consisted of the plasma only, the beam representing the perturbation, then this parameter would have to remain much smaller than 1 within the framework of a linear response theory. In turns out that the dispersion equation which has just been derived is the dispersion equation of the beam+plasma system. The perturbed ground state is therefore the sum of the beam and the plasma. It is thus perfectly possible to investigate the linear response of the whole system even when $\alpha=1$ so that we do need to make any assumption regarding this parameter.

\section{Classical magnetized plasma}
Before we turn to the quantum case, let us quickly remind some basic features of the cold magnetized filamentation instabilities \cite{Godfrey1975} in the classical (non-quantum) regime. To this extent, the dispersion equation, which is just the determinant of the tensor we just defined, is solved numerically and Figure \ref{fig:1} displays the growth rates obtained with and without the magnetic field (curves a and b). The stabilizing effect of the magnetic field is twofold. On one hand, the smallest unstable wave vector switches from $Z=0$ to
\begin{equation}\label{eq:ZsmallClass}
   Z_1=\frac{\beta\Omega_b\sqrt{1+\alpha}}{\sqrt{\alpha(1+\alpha)\beta^2-\Omega_B^2}}.
\end{equation}
On the other hand, the growth rate saturation value $\delta_\infty$ for large $Z$ is lower with
\begin{equation}\label{eq:TauxClassGdZ}
    \delta_\infty=\sqrt{\alpha(1+\alpha)\beta^2-\Omega_B^2},
\end{equation}
which vanishes exactly for
\begin{equation}\label{eq:BClass}
\Omega_B=\Omega_{Bc}\equiv\beta\sqrt{\alpha(1+\alpha)}
\end{equation}
Noteworthily, this value of the magnetic field also makes the quantity $Z_1$ diverge. The physical interpretation of this threshold is simple as $\beta\sqrt{\alpha(1+\alpha)}$ is just the maximum growth rate of the instability in the non-magnetized case \cite{BretPoPHierarchie}. Filamentation instability is thus inhibited when the electron response to the magnetic field is quicker.

\section{Quantum magnetized plasma}
\subsection{Marginal stability analysis}
Figure \ref{fig:1}c displays the growth rate in terms of $Z$ accounting for quantum effects. As in the non-magnetized case \cite{BretPoPQuantum2007}, quantum effects introduce a cut-off at large $Z$ so that we now have to characteristic wave vectors $Z_1$ and $Z_2$ determining the instability range. Both of them can be investigated directly from the dispersion equation. Since the growth rate vanishes for these wave vector while the root yielding the filamentation instability has no real part, we can write
\begin{equation}
    \det\mathbf{T}(x=0,Z=Z_{1,2})=0.
\end{equation}
It turns out that this equation can be simplified. After replacing $Z^2\rightarrow \mathcal{Z}$ and eliminating $x=0$ as a double root of the dispersion equation, we find that the equation above is equivalent to $P(\mathcal{Z})Q(\mathcal{Z})=0$ with
\begin{eqnarray}\label{eq:P(Z)}
    P(\mathcal{Z})&=&(\Theta_c\mathcal{Z}^2+\beta^4\Omega_B^2)(\mathcal{Z}+(1+\alpha)\beta^2)-\mathcal{Z}\alpha(1+\alpha)\beta^6\nonumber\\
    Q(\mathcal{Z})&=&\mathcal{Z}\beta^4\Omega_B^2+(\mathcal{Z}+(1+\alpha)\beta^2)(\Theta_c\mathcal{Z}^2+(1+\alpha)\beta^4).
\end{eqnarray}

\begin{figure}[t]
\begin{center}
\includegraphics[width=0.45\textwidth]{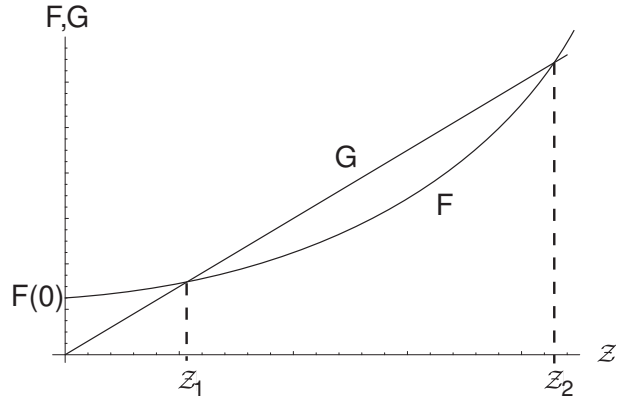}
\end{center}
\caption{Schematic representation of the functions $F$ and $G$ defined by Eqs. (\ref{eq:FGraph}).} \label{fig:2}
\end{figure}

Every term of the second polynomial is clearly positive so that it yields only negative roots $\mathcal{Z}<0$, implying some complex wave vector. Because we seek real wave vectors, we can conclude that $\mathcal{Z}_1=(Z_1)^2$ and $\mathcal{Z}_2=(Z_2)^2$ are both zero's of $P(\mathcal{Z})$. This function being a polynomial of the third order, it is possible to find the exact solutions. We nevertheless use some graphical method for a more intuitive approach. Let us then define
\begin{eqnarray}\label{eq:FGraph}
    F(\mathcal{Z})&=&(\Theta_c\mathcal{Z}^2+\beta^4\Omega_B^2)(\mathcal{Z}+(1+\alpha)\beta^2),\nonumber\\
    G(\mathcal{Z})&=&\mathcal{Z}\alpha(1+\alpha)\beta^6,
\end{eqnarray}
so that $P=0$ is equivalent to $F=G$. $F$ is a third order polynomial, monotonically increasing for $\mathcal{Z}>0$, and starting from $F(0)= (1+\alpha)\Omega_B^2\beta^6$ with an initial slope $F'(0)= \Omega_B^2\beta^4$. $G$ is a first order monotonically increasing polynomial with $G(0)=0$ and slope $\alpha(1+\alpha)\beta^6$. We can now conduct the graphical analysis of the problem following the guidelines set by the schematic representation of $F$ and $G$ on Figure \ref{fig:2}. When increasing $\Omega_B$ or $\Theta_c$, the curve $G$ is not modified because neither $\Omega_B$ nor $\Theta_c$ appear in its expression. Meanwhile, $F(0)$ increases with $\Omega_B$, and $F(\mathcal{Z})$ increases all the more than $\Theta_c$ and $\Omega_B$  are large. This allows us to draw the following conclusions:
\begin{itemize}
  \item In the absence of magnetic field, $F(0)=F'(0)=0$ while the previous analysis remains unchanged. The equation $F=G$ thus has two positive solutions regardless of the other parameters. One solution is $\mathcal{Z}_1=0$, i.e. $Z=0$, and we label the other $\mathcal{Z}_2>0$. We recover the existence of a quantum cut-off \cite{BretPoPQuantum2007} at large wave vector, and prove here that the instability is never completely stabilized since $\mathcal{Z}_2$ never vanishes.
  \item For any finite magnetic field, one has $F(0)>0$ and $F'(0)>0$, and the typical resulting situation is the one represented on Fig. \ref{fig:2}. As long as $F(0)$, or the growth of $F(\mathcal{Z})$, are ``not too high'', the equation $F=G$ has two positive roots $\mathcal{Z}_{1,2}=(Z_{1,2})^2$. But it is obvious that as $\Omega_B$ or $\Theta_c$ increase, the too real roots become one before they vanish. We thus come to conclusion that the instability can be completely suppressed by quantum effects if, and only if, the system is magnetized, regardless of the strength of the magnetic field. It is graphically obvious that since an increase of both $\Omega_B$ or $\Theta_c$ contribute to the collapse of the two reals roots, the stabilization condition should result in a balance between these quantities. It should be possible to stabilize the system at low $\Omega_B$ with an high $\Theta_c$, or vice versa.
\end{itemize}

\begin{figure*}[t]
\begin{center}
\includegraphics[width=0.9\textwidth]{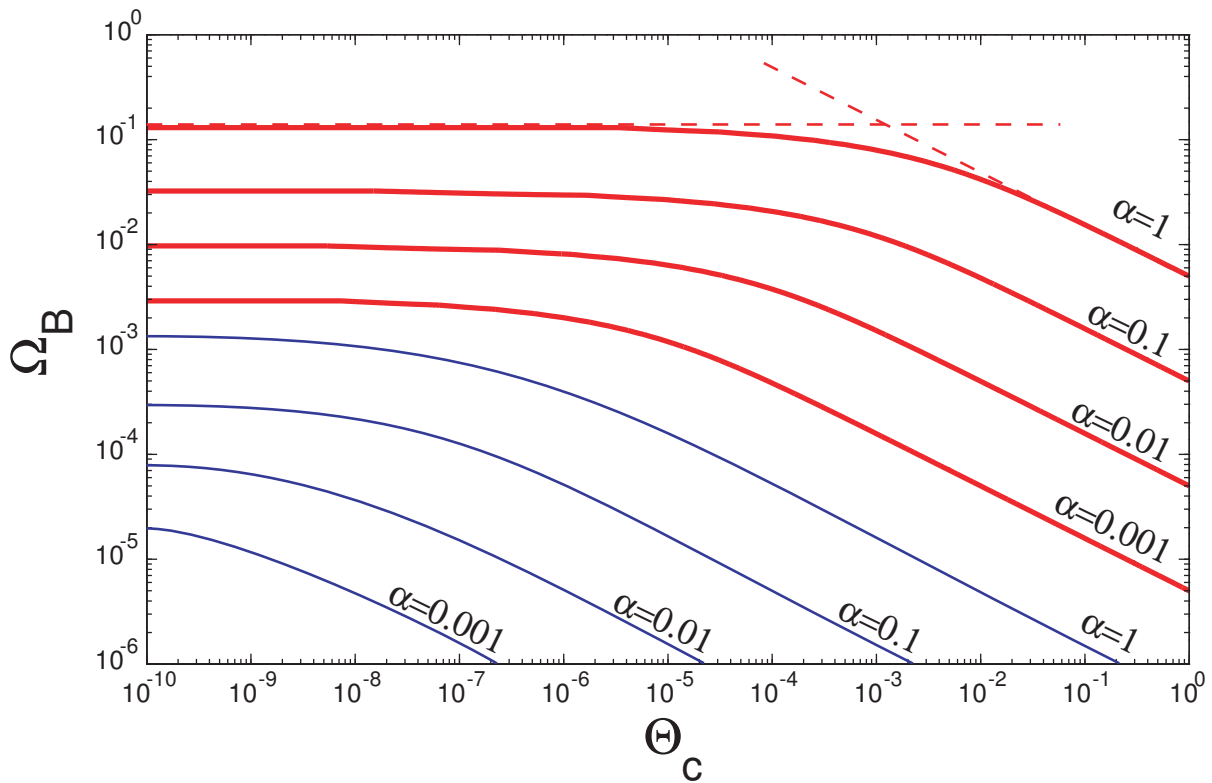}
\end{center}
\caption{(Color online) Values of $\Theta_c^*$ and $\Omega_B^*$ implicitly defined by Eqs. (\ref{eq:ZL},\ref{eq:relation}) for various beam to plasma density ratios $\alpha$ and $\beta=0.1$ (red bold curves) and $\beta=10^{-3}$ (blue thin curves). Given $\alpha$ and $\beta$, the system is marginally stable for parameters  ($\Theta_c^*,\Omega_B^*$) located on the corresponding curve, and stable above. The value of $\Omega_{Bc}$ (see Eq. \ref{eq:BClass}) for $\beta=0.1$ and $\alpha=1$ is represented by the horizontal dashed red line, and the oblique dashed curve corresponds to Eq. (\ref{eq:OmegaHighTheta}) with the same $\alpha,\beta$.} \label{fig:3}
\end{figure*}

In the magnetized case, the instability is marginal when the equation $F=G$ has one double root $\mathcal{Z}_1=\mathcal{Z}_2=\mathcal{Z}_L$. Here, $L$ stands for $L$ast because $Z_L=\sqrt{\mathcal{Z}_L}$ is eventually the last unstable wave vector before complete stabilization. If $\mathcal{Z}_L$ is double root of $F$ for the parameters defining the marginal stability, then for these very parameters $F$ can be cast under the form $F(\mathcal{Z})=(\mathcal{Z}-a)(\mathcal{Z}-\mathcal{Z}_L)^2$ where $a$ is the third root. By developing this last form and identifying the coefficients of the polynomial with the ones extracted from Eq. (\ref{eq:FGraph}), we can write the following equations,
\begin{eqnarray}\label{eq:coefs}
  a+2\mathcal{Z}_L &=& -(1+\alpha)\beta^2, \\
  2a\mathcal{Z}_L+\mathcal{Z}_L^2 &=& \frac{\beta^4\Omega_B^{*2}-\alpha(1+\alpha)\beta^6}{\Theta_c^*}, \\
  a\mathcal{Z}_L^2 &=& -\frac{(1+\alpha)\beta^6\Omega_B^{*2}}{\Theta_c^*},
\end{eqnarray}
where the superscript * refers to the values at marginal stability. By eliminating $a$ between the first and the second equation, one finds a second order equation for $\mathcal{Z}_L$ which positive solution can be cast under the form,
\begin{equation}\label{eq:ZL}
 \mathcal{Z}_L=\frac{\beta^2(1+\alpha)}{3}\left(\sqrt{1+3\frac{\Omega_{Bc}^2-\Omega_{B}^{*2}}{\Theta_c^*(1+\alpha)^2}}-1\right).
\end{equation}
Then, eliminating $a$ between the first and the third yields an implicit relation between $\mathcal{Z}_L$, $\Theta_c^*$ and $\Omega_B^*$ at marginal stability,
\begin{equation}\label{eq:relation}
   (1+\alpha)\beta^6\Omega_B^{*2}=\mathcal{Z}_L^2\Theta_c^*(2\mathcal{Z}_L+(1+\alpha)\beta^2).
\end{equation}
Equations (\ref{eq:ZL},\ref{eq:relation}) therefore define  $\Theta_c^*$ and $\Omega_B^*$ in terms of each other, and of the others parameters of the problem. The curves thus defined appear on Figure \ref{fig:3} for various $\alpha$'s and $\beta$'s. Parameters ($\Theta_c,\Omega_B$) located above a given curve ($\Theta_c^*,\Omega_B^*$) define a completely stabilized system.

\section{Analytical expressions for marginal stability}
\subsection{Classical limit}
We observe on Fig. \ref{fig:3} that $\Omega_B^*$ reaches a finite value when $\Theta_c^*\rightarrow 0$. This classical limit is obviously the marginal magnetic parameter  $\Omega_{Bc}$ given by Eq. (\ref{eq:BClass}). We thus assume a leading term in the development of $\Omega_B^*$ for small  $\Theta_c^*$  of the form $\Omega_B^* = (1-\kappa\Theta_c^{*\xi})\Omega_{Bc}$. Inserting this expression in Eqs. (\ref{eq:ZL},\ref{eq:relation}) and expanding the results in series of $\Theta_c^*$, we find
\begin{equation}\label{eq:Omega0}
    \Omega_B^*(\Theta_c^*\rightarrow 0) \sim \left(1-\frac{3(1+\alpha)^{1/3}}{2^{5/3}\alpha^{1/3}\beta^{2/3}}\Theta_c^{*1/3}\right)\Omega_{Bc},
\end{equation}
and
\begin{equation}\label{eq:ZLAsympt}
    \mathcal{Z}_L(\Theta_c^*\rightarrow 0) \sim \frac{\alpha^{1/3}(1+\alpha)^{2/3}\beta^{8/3}}{2^{1/3}\Theta_c^{*1/3}}.
\end{equation}
In accordance with the classical case where the smallest unstable wave vector diverges for marginal stability (see Eqs. \ref{eq:ZsmallClass},\ref{eq:TauxClassGdZ}), the last unstable wave vector $Z_L$ behaves like $1/\Theta_c^{*1/6}$ in the weak quantum regime since $\mathcal{Z}_L=Z_L^2$.

\subsection{Strong quantum limit}
Having elucidated the weak quantum regime, we now turn to the strong quantum one. Figure \ref{fig:3} makes it clear that marginal stability behaves differently within each regime. In order to discuss this point, let us consider expression (\ref{eq:ZL})  of $\mathcal{Z}_L$ in terms of the marginal classical magnetic parameter $\Omega_{Bc}$. In the ``large'' $\Theta_c^*$ regime, the ratio under the square root becomes small compared to unity, and Fig. \ref{fig:3} shows that $\Omega_{B}^*\ll \Omega_{Bc}$. Developing the square root, we find directly
\begin{equation}\label{eq:ZL3}
 \mathcal{Z}_L=\frac{\beta^2\Omega_{Bc}^2}{2\Theta_c^*(1+\alpha)}.
\end{equation}
In this strongly quantum regime, the last unstable wave vector thus tends to zero like $1/\sqrt{\Theta_c^*}$. Inserting the former expression in Eq. (\ref{eq:relation}) yields the magnetic parameter required to stabilize the system
\begin{equation}\label{eq:OmegaHighTheta}
    \Omega_B^*\sim \frac{\Omega_{Bc}^2}{2(1+\alpha)\Theta_c^*}=\frac{\alpha\beta^2}{2\sqrt{\Theta_c^*}}.
\end{equation}
This limit is plotted on Fig. \ref{fig:3} for $\beta=0.1$ and $\alpha=1$ and perfectly fits the numerical evaluation for large $\Theta_c^*$.

It is now possible to exhibit the dimensionless parameter measuring the strength of quantum effects. The equation above indicates that $\Omega_B^*\ll\Omega_{Bc}$ if $\Omega_{Bc}\ll 2(1+\alpha)\Theta_c^*$, and the curves plotted on Fig. \ref{fig:3} demonstrate that a reduction of the stabilizing parameter is the signature of the strong quantum regime. Because we think here in terms of orders of magnitudes, we drop the $2(1+\alpha)$ factor and finally define
\begin{equation}\label{eq:Xi}
    \Lambda=\frac{\Omega_{Bc}}{\Theta_c},
\end{equation}
as the parameter determining the strength of quantum effects. These are weak for $\Lambda\gg 1$ and strong in the opposite limit $\Lambda\ll 1$.

\section{Unstable systems}
Having elucidated  how the system can be completely stabilized by quantum magnetic effects, we now turn to unstable systems in order to investigate the growth rate of the instability and the most unstable wave vector for a given configuration.

\begin{figure}[t]
\begin{center}
\includegraphics[width=0.45\textwidth]{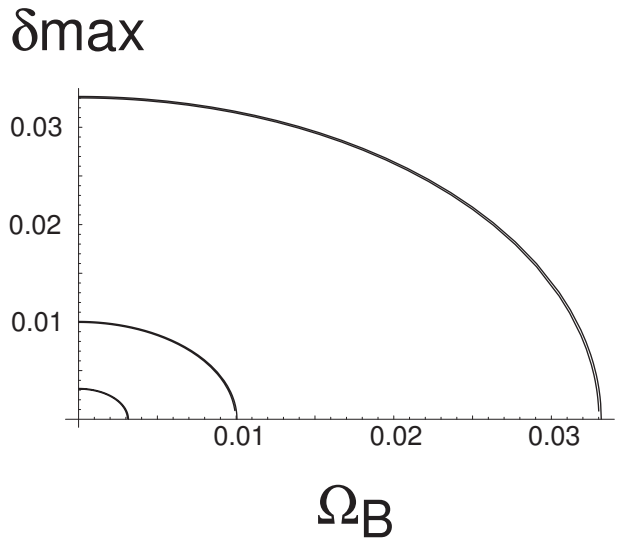}
\end{center}
\caption{Maximum growth rates in terms of $\Omega_B$ in the classical and quantum cases for $\beta=0.1$ and $\alpha=10^{-1},10^{-2},10^{-3}$. With $\Theta_c=10^{-10}$ ($n_p=7.6\times 10^{22}$ cm$^{-3}$), the parameter $\Lambda$ given by Eq. (\ref{eq:Xi}) is always larger than $3\times 10^7$ and the classical and quantum curves  are hardly distinguishable although the quantum growth rate is a little bit smaller.} \label{fig:4}
\end{figure}

\subsection{Maximum growth rate}
In the weakly quantum regime with $\Lambda\gg 1$, the maximum quantum growth rate is very close to its classical counterpart all the way down to complete stabilization which, as we just mentioned, occurs for similar magnetic parameters $\Omega_B$ (see Eq. \ref{eq:Omega0} above). Figure \ref{fig:4} present a plot of the maximum growth rates along the $Z$ axis, in terms of $\Omega_B$ in the classical and quantum cases for $\beta=0.1$ and various $\alpha$'s. Parameters have been chosen to illustrate the present weak quantum regime. Such a system can thus be viewed as basically magnetized with some weak quantum effects, and stabilization mainly comes from the magnetic field.

\begin{figure}[t]
\begin{center}
\includegraphics[width=0.8\textwidth]{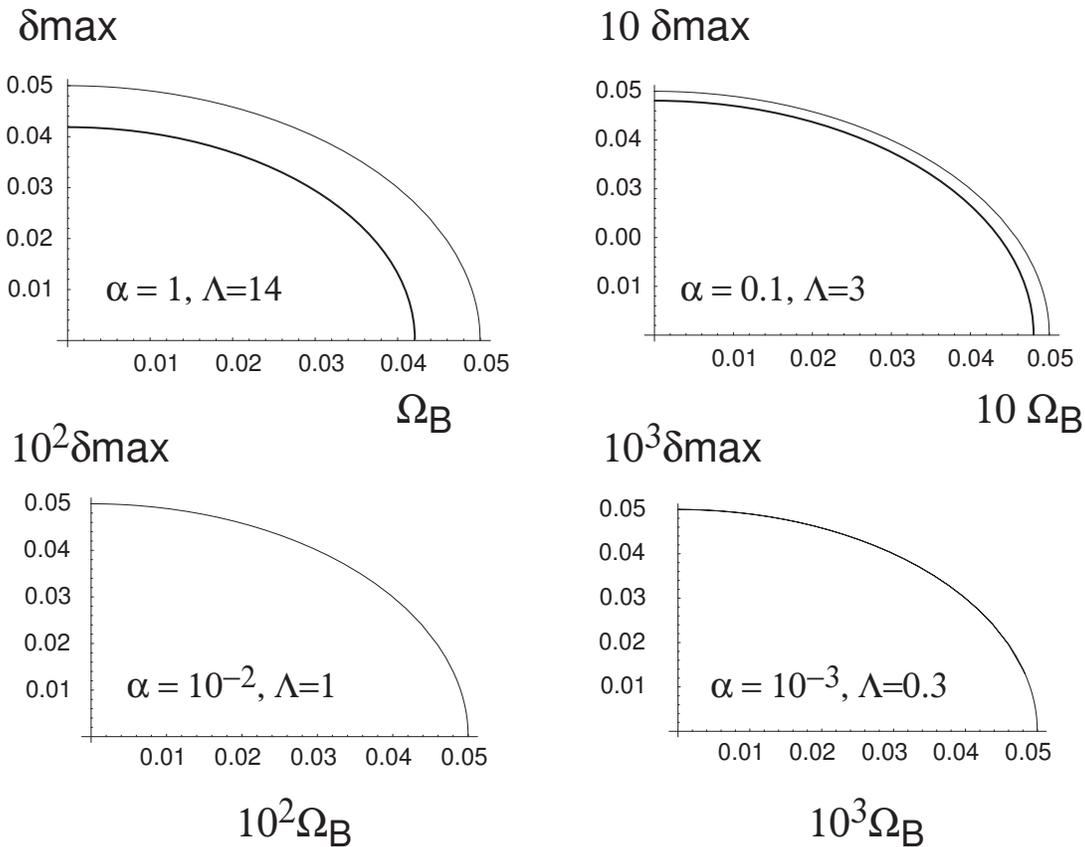}
\end{center}
\caption{Maximum growth rates in terms of $\Omega_B$ for $\beta=0.1$ and $\Theta_c=10^{-2}$. The thin curves have been computed numerically, and the bold ones (when distinguishable from the thin) correspond to Eq. (\ref{eq:dmaxB}). Agreement improves with $\Lambda$ decreasing.} \label{fig:5}
\end{figure}

When $\Lambda\ll 1$ (strong quantum regime), stabilization is reached earlier with respect to $\Omega_B$ (see Fig. \ref{fig:3} and Eq. \ref{eq:OmegaHighTheta}). Here, stabilization comes from a combination of quantum and magnetic effects, as indicated by the oblique slope of the curves in Fig. \ref{fig:3}. We plot on Figure \ref{fig:5} the maximum growth rate  in terms of $\Omega_B$. We recognize the kind of curve obtained for a classical magnetized plasma $\delta_{max}^2=\Omega_B^{\mathrm{cut~off}~2}-\Omega_B^2$ with a ``cut-off'' magnetic parameter $\alpha\beta^2/2\sqrt{\Theta_c^*}$. This is why we plotted together the numerical evaluation of the maximum growth rate together with the function,
\begin{equation}\label{eq:dmaxB}
    \delta_{max}=\sqrt{\left(\frac{\alpha\beta^2}{2\sqrt{\Theta_c^*}}\right)^2-\Omega_B^2}.
\end{equation}
It can be checked that this function fits the result all the more than $\Lambda$ is small. With Eq. (\ref{eq:OmegaHighTheta}), we then come to the conclusion that as far as the maximum growth rate is concerned, strong quantum effects are equivalent to the substitution,
\begin{equation}\label{eq:Beq}
    \Omega_{Bc}\Leftrightarrow \Lambda\frac{\Omega_{Bc}}{2(1+\alpha)}.
\end{equation}
Because this new quantum cut-off is much smaller than the classical one, the maximum growth rate is reduced accordingly.

\subsection{Most unstable wave vector}
The most unstable wave vector is, together with the maximum growth rate, the most relevant information about the unstable system. In the classical case, the growth rate just saturates at large $Z$ yielding a continuum of most unstable modes. But quantum effects stabilize the large $Z$ modes, so that there is always one mode growing faster that the others.

For systems near marginal stability, the last unstable wave vector $Z_L=\sqrt{\mathcal{Z}_L}$, given exactly by Eq. (\ref{eq:ZL}), and in the weak and strong quantum limits by Eqs. (\ref{eq:ZLAsympt},\ref{eq:ZL3}) respectively, is by definition a very good approximation of this most unstable wave vector when replacing the marginal parameter $\Theta_c^*$ by its actual value $\Theta_c$. Indeed, we found numerically that expressions (\ref{eq:ZLAsympt},\ref{eq:ZL3}) are still quite accurate, even for systems far from stabilization. This can be understood  from Fig. \ref{fig:2}: on one hand, the most unstable wave vector for a given configuration is necessarily between $Z_1$ and $Z_2$. On the other hand, the last unstable wave vector belongs to the same interval because $Z_1$ increases while $Z_2$ decreases as the system moves towards stabilization. For a typical situation such as the one represented on Fig. \ref{fig:1}c, $Z_1$ and $Z_2$ are eventually quite close to each other so that $Z_L$, which is in between, cannot be far from the most unstable wave vector. With the parameters chosen for this plot, we find $\Lambda=2.5\times 10^5$ indicating a weak quantum regime. We therefore turn to Eq. (\ref{eq:ZLAsympt}) and find the most unstable wave vector for $Z\sim 0.4$, which fits accurately what is observed.

\section{Discussion}
Quantum effects have been assessed with respect to the filamentation instability in a magnetized plasma. As far as the unstable wave vector range is concerned, magnetic effects set it a finite lower bound, while quantum effects introduce a cut-off at large $k$. As a result, the unstable domain takes the form $[k_1,k_2]$ and can eventually vanish for some parameters configurations which were elucidated.

We also found that the dimensionless parameter $\Lambda=\Omega_{Bc}/\Theta_c$ determines the strength of quantum effects. When $\Lambda\gg 1$, the instability can be described in classical terms, and eventually vanishes when increasing the magnetic field, while the unstable wave vector range shifts towards infinity. When quantum effects are strong, namely $\Lambda\ll 1$, the instability still vanishes with the magnetic field, but the unstable wave vector range tends to zero. Furthermore, the magnetic field required to stabilize the system is divided by $\Lambda/2(1+\alpha)\ll 1$ with respect to its classical value, so that filamentation can be suppressed by a much smaller magnetic field than in the non-quantum case. These results may have important consequences when dealing with dense space plasmas.

Finally, it will be necessary to assess both relativistic effects, which tend to enlarge the instability domain while reducing the maximum growth rate \cite{BretPRE2004}, and kinetic effects which usually have a stabilizing effect \cite{BretPRE2005}. To this extent, relativistic quantum kinetic theory will be required, or the relativistic form of the quantum Euler equation (\ref{eq:force}) will have to be elaborated. As long as the theory implemented is non-relativistic, the magnetic stabilization level unraveled here should remain an \emph{upper} stabilization bounds when kinetic effects are accounted for. In the classical relativistic regime, it has been demonstrated that the stabilizing magnetic field behaves like $\sqrt{\gamma_b}$  \cite{Cary1981}, where $\gamma_b$ is the relativistic factor of the beam. Because this increase of the magnetic threshold eventually stems from the relativistic increase of the mass of the electrons, we can conjecture that the same factor $\sqrt{\gamma_b}$ will be found in the relativistic counterpart of Eq. (\ref{eq:OmegaHighTheta}), but this shall need confirmation.

\section{Acknowledgements}
This work has been  achieved under projects FIS 2006-05389 of the
Spanish Ministerio de Educaci\'{o}n y Ciencia and PAI-05-045 of
the Consejer\'{i}a de Educaci\'{o}n y Ciencia de la Junta de
Comunidades de Castilla-La Mancha. Thanks are due to Laurent Gremillet for enriching discussions.

\end{document}